\documentclass[aps,prx,twocolumn,superscriptaddress,floats,floatfix,notitlepage,longbibliography]{revtex4-1}
\usepackage{amssymb,amsmath,amsthm,amssymb,mathrsfs}
\usepackage{graphicx}
\usepackage{color}
\usepackage{epstopdf}
\usepackage{epsfig}
\usepackage{rotating}
\usepackage{float}
\usepackage{ulem,bm}
\usepackage{xr}
\usepackage{hyperref}
\usepackage{breakurl}
\usepackage{lipsum}
\usepackage{footnote}

\newcommand{\REM}[1]{{}}

\newcommand{\df}{\displaystyle\frac}
\begin{document}

\title{Ephemeral Antibubbles: Spatiotemporal Evolution from Direct
Numerical Simulations}

\author{Nairita Pal\footnote{nairitap2009@gmail.com}}
\affiliation{Los Alamos National Laboratory, NM 87545, USA}
\author{Rashmi Ramadugu\footnote{rashmi.ramadugu@gmail.com}}
\affiliation{TIFR Center for Interdisciplinary Sciences, Tata Institute of Fundamental Research, Gopanpally, Hyderabad, 500046, India}
\author{Prasad Perlekar\footnote{perlekar@tifrh.res.in}}
\affiliation{TIFR Center for Interdisciplinary Sciences, Tata Institute of Fundamental Research, Gopanpally, Hyderabad, 500046, India}
\author{Rahul Pandit\footnote{rahul@iisc.ac.in; also at Jawaharlal Nehru Centre For
Advanced Scientific Research, Jakkur, Bangalore, India.}}
\affiliation{Centre for Condensed Matter Theory, Department of Physics, Indian Institute 
of Science, Bangalore 560012, India}

\date{\today}
\begin{abstract}

Antibubbles, which consist of a shell of a low-density fluid inside a
high-density fluid, have several promising applications. We show, via
extensive direct numerical simulations (DNSs), in both two and three
dimensions (2D and 3D), that the spatiotemporal evolution of
antibubbles can be described naturally by the coupled
Cahn-Hilliard-Navier-Stokes (CHNS) equations for a binary fluid. Our DNSs
capture elegantly the gravity-induced thinning and breakup of an
antibubble via the time evolution of the Cahn-Hilliard scalar order
parameter field $\phi$, which varies continuously across interfaces, so
we do not have to enforce complicated boundary conditions at the moving
antibubble interfaces. To ensure that our results are
robust, we supplement our CHNS simulations with
sharp-interface  Volume-of-Fluid (VoF) DNSs. We track the thickness of the
antibubble and calculate the dependence of the lifetime of an
antibubble on several parameters; we show that our DNS results agree
with various experimental results; in particular, the velocity with
which the arms of the antibubble retract after breakup scales as
$\sigma^{1/2}$, where $\sigma$ is the surface tension.

\end{abstract}

\pacs{47.27.E-,47.27.eb,47.55.P-}
\maketitle


\section{Introduction}
\label{sec:Introduction}

Antibubbles, which comprise a shell of a low-density fluid inside a
high-density fluid, have been known for close to $90$ years, since the work of
Hughes and Hughes~\cite{hughes1932liquid}.  In contrast to an ordinary bubble,
an antibubble has two surfaces, which trap a certain volume of fluid between
them.  Therefore, the contact area of an antibubble is much larger than that of
a bubble with the same fluid volume; this property can be exploited for a
variety of chemical reactions. Furthermore, antibubbles are commonly used in
clinical diagnostic imaging, sonoporation (see, e.g.,
Ref.~\cite{kotopoulis2014sonoporation}), as agents for ultrasound-guided drug
delivery~\cite{johansen2015nonlinear,
kotopoulisa2015acoustically,postema2007ultrasound}, and for active leakage
detection~\cite{johansen2015ultrasonically}. Clearly, an antibubble is unstable
in the presence of gravity, and, if the inner core of the antibubble is denser than
the outer core, the antibubble rises under gravity; because of hydrostatic
pressure in the outer core, the fluid rises from the bottom to the top,
resulting in a thinning, and subsequent collapse, of the outer shell.  Although
there have been a number of experimental investigations of the spatiotemporal
evolution of an antibubble~\cite{dorbolo2006vita,
kim2006antibubbles,dorbolo2010antibubble,dorbolo2003fluid,bai2016formation,kalelkar2017inveterate,
chatzigiannakis2021thin,chatzigiannakis2020breakup,song2020criteria} and drops~\cite{will2021kinematics,li2021marangoni,soto2020ultrasound,soto2020diffusive,prakash2016energy} to name a few,
over the past few decades, theoretical studies of antibubble evolution have been initiated
only recently~\cite{scheid2012antibubble,zou2013collapse,
sob2015theory,yang2020mathematical,vitry2019controlling,song2020criteria}
and they have not, attempted, hitherto, to address the spatiotemporal
evolution of antibubbles in detail. The number of experimental studies of the
complete spatiotemporal evolution of antibubbles is also limited, partly because
great care has to be exercised to stabilise antibubbles; often, surfactant
molecules have to be introduced into the high-density liquid phase for such
stabilization.

We develop a \textit{natural, multiphase} model for antibubbles and
demonstrate how we can use direct numerical simulations (DNSs) to follow the
spatiotemporal development of ephemeral, but beautiful, antibubbles. We show
that the Cahn-Hilliard-Navier-Stokes (CHNS) equations, which have been used to
study a variety of problems in binary-fluid
flows~\cite{boffetta2012two,perlekar2014spinodal,scarbolo2013unified,
pal2016binary, perlekar2017two,gibbon2016regularity,gibbon2018BKM,
fan2016cascades,fan2017formation, fan2018chns}, provide a \textit{minimal
theoretical framework} for studying the spatiotemporal evolution of
antibubbles; in addition to a velocity field, the CHNS system employs a phase
field $\phi$ that distinguishes between the two fluid phases.
We use the CHNS equations~\cite{boffetta2012two,
perlekar2014spinodal, pal2016binary, gibbon2016regularity, perlekar2017two,
gibbon2018BKM} to study antibubbles in two-dimensions (2D) and in three
dimensions (3D) by using extensive DNSs. To complement our CHNS results, we
also employ an alternative Volume-of-Fluid (VoF) numerical scheme that is a
sharp-interface method.


Our studies yield several interesting results that (a) provide explanations for
many experimental observations and (b) suggest new experimental studies: Our
results for the spatiotemporal evolution of antibubbles, in 2D and 3D, show
clearly how antibubble breakup occurs either because of gravity-induced
thinning or the puncturing of its bottom boundary; the the collapsing
antibubble then forms a rim that retracts with a velocity $v_{rim}$. We uncover
signatures of this collapse in Fourier-space spectra of the Cahn-Hilliard field
$\phi$, the velocity, and the vorticity. We show that $v_{rim} \sim
\sigma^{1/2}$, where $\sigma$ is the surface tension; and this power-law
exponent is independent of the kinematic viscosity $\nu$ of the background
fluid.  We investigate the dependence of the scaled antibubble lifetime
$\tau_1/\tau_g$ on $\sigma$, $\nu$, and the scaled outer radius of the
antibubble $R_0/h_0$, in 2D; here, $\tau_g\equiv \sqrt{R_0/Ag}$, with $A$ the
Atwood number and $g$ the acceleration due to gravity, and $h_0$ is the initial
thickness of the antibubble shell. We compare our results with experiments and
earlier theoretical studies. 

The remainder of this paper is organized as follows: In Sec.~\ref{sec:Model} we
present the models and the numerical methods that we use. In
Sec.~\ref{sec:Results} we present the results our DNSs in the following
subsections: 
\begin{itemize}
\item subsection~\ref{subsec:SpTem2D} on the spatiotemporal evolution of antibubbles 
(2D DNSs); 
\item subsection~\ref{subsec:Spectra2D} on the temporal evolution of Fourier spectra 
of $\phi$ and the velocity and vorticity fields that are associated with
antibubbles in our 2D DNSs; 
\item subsection~\ref{subsec:Rim} on the velocity of the retracting rim of the ruptured 
antibubble in our 2D DNSs;
\item subsection~\ref{subsec:EnTime} on the signatures of this rupture in the time 
dependence of the energy; 
\item subsection~\ref{subsec:Size} on the dependence of the antibubble-rupture time on 
the size of the antibubble and the surface tension; 
\item subsection~\ref{subsec:Viscosity} on the dependence of the antibubble-rupture 
time on the kinematic viscosity; 
\item subsection~\ref{subsec:SpTem3D} on the spatiotemporal evolution of antibubbles
in our 3D DNSs.
\end{itemize}	
We present conclusions and a discussion of our results in Sec.~\ref{sec:Conclusions}.

\section{Model and Numerical Methods}
\label{sec:Model}

An antibubble consists of a shell of light fluid of density $\rho_1$ inside a
background heavy fluid of density $\rho_2$; these fluids are immiscible and
incompressible. The following CHNS equations provide a natural theoretical
description for such a binary-fluid
mixture~\cite{celani2009phase,scarbolo2013turbulence,yue2004diffuse,scarbolo2011phase}:

\begin{align}
\rho(\phi) D_t {\bm u} &= -\nabla p +
       \varsigma(\phi) \nabla^2 {\bm u} + \mathbf{F}_{\sigma}  \nonumber\\
	&+[\rho(\phi)- \displaystyle\frac{\rho_1+\rho_2}{2}]{\bm g} - \alpha \mathbf{u};
\label{ns}\\
D_t \phi  &=  \gamma  \nabla^2 {\mu}~{\rm{and}}~ \nabla \cdot {\bm
u} =0; \label{ch}
\end{align}
In Eqs.\eqref{ns}-\eqref{ch}, $D_t \equiv \left(\partial_t + {\bm u}\cdot
\nabla\right)$ is the convective derivative, ${\bm u}\equiv(u_x,u_y,u_z)$ is
the fluid velocity, $\alpha$ is the coefficient of friction (we use this only
in 2D), ${\bm g}$ is the constant acceleration due to gravity, which points
downwards, $\rho(\phi)\equiv \rho_1(1+\phi)/2 + \rho_2 (1-\phi)/2$ is the
density, $\varsigma(\phi)\equiv \nu_1(1+\phi)/2 + \nu_2 (1-\phi)/2$ is the
dynamic viscosity, $\phi({\bm x},t)$ is the order-parameter field at the point
${\bm x}$ and time $t$ [with $\phi({\bm x},t)>0$ in the background (majority)
phase and $\phi({\bm x},t)<0$ in the antibubble-shell (minority) phase].  For the
illustrative results that we present, we choose $\rho(\phi)$ and
$\varsigma(\phi)$ such that the kinematic viscosity $\nu$ is identical in the
two phases. 

When we study 2D flows, we use the following stream-function-vorticity 
formulation~\cite{boffetta2012two,pandit2009statistical}: 
\begin{eqnarray}
 \left(\partial_t + {\bm u}\cdot \nabla\right) \omega &=& \nu \nabla^2 \omega - 
\alpha \omega-\nabla \times (\phi \nabla \mu) - A \nabla\phi \times {\bm g}; \nonumber \\
 \label{ns_2d}
 \left(\partial_t + {\bm u}\cdot \nabla\right){\bf \phi} & = & \gamma \nabla^2 {\mu}~{\rm{and}}~ \nabla \cdot {\bm u} =0.
 \label{ch_2d}
 \end{eqnarray}
Here, $\omega=(\nabla \times {\bm u}) {\hat{\bm e}}_z$ is the vorticity.
We obtain ${\omega}({\bm x},t)$ and $\phi({\bm x},t)$
from our 2D DNS~\cite{supmat}, and from these we calculate the total kinetic
energy $E(t)=\langle |{\bm u}({\bm x},t)|^{2} \rangle_{{\bm x}}$ and the
fluid-energy dissipation rate $\varepsilon(t)=\langle \nu |\omega({\bm
x},t)|^{2} \rangle_{{\bm x}}$, where $\langle \rangle_{{\bm x}}$ denotes the
average over space, and $\omega=\nabla \times {\bm u}$ is the fluid vorticity.
We calculate the lifetime of an antibubble falling under
gravity in two ways: (a) from the energy time series; and (b) from the minimum
thickness at the South pole of the antibubble.  

At time $t=0$ we begin with the order-parameter profile~\cite{celani2009phase}
\begin{eqnarray}
\phi({\bm x}) &=&-\tanh\left[\frac{1}{\sqrt{2}\xi}\left(|{\bm x}-{\bm x}_c|- R_0\right)\right]\nonumber \\
&&+\tanh\left[\frac{1}{\sqrt{2}\xi}\left(|{\bm x}-{\bm x}_c|-R_1\right)\right]-1.0,
\label{tanh}
\end{eqnarray}
where $R_0$ and $R_1$ are, respectively, the initial magnitudes of the outer
and inner radii of the antibubble, whose centre is initially at ${\bm x}_c$;
the interface width $\xi$ is measured by the dimensionless Cahn number
$Ch=\xi/L$. We use the initial thickness of the antibubble shell $h_0 \equiv R_0-R_1$
as a typical length scale. 

In our CHNS studies we use the Boussinesq approximation, wherein  $\rho(\phi)D_t
{\bm u} \approx \displaystyle\frac{\rho_1+\rho_2}{2} D_t {\bm u}$ and
$[\rho(\phi)- \displaystyle\frac{\rho_1+\rho_2}{2}]{\bm g}\approx -A \phi g$ in
Eq.\eqref{ns}; the surface-tension force $\mathbf{F}_{\sigma}\equiv -\phi
\pmb{\nabla}\mu$ and the chemical potential $\mu({\bm x},t)$ follow from the
Cahn-Hilliard free-energy functional ${\mathcal F}$ (see the Supplemental
Material~\cite{supmat}):
\begin{eqnarray} 
\mu & = & \delta {\mathcal F}[\phi]/\delta \phi({\bm x},t); \nonumber \\
{\mathcal F}[\phi] & = & \Lambda \int [(\phi^2-1)^2/(4\xi^2) + |\nabla
\phi|^2/2] d{\bm x};  
\label{eq:Free-en}
\end{eqnarray} 
%
%
here, $\Lambda$ is the energy density with which the two phases mix in the
interfacial regime~\cite{celani2009phase}, $\xi$ sets the scale of the
interface width , $\sigma=\frac{2\sqrt{2}\Lambda}{3\xi}$ is the surface
tension, $\gamma$ is the mobility~\cite{yue2004diffuse} of the binary-fluid
mixture; the Cahn number $Ch=\xi/L$, where $L$ is the linear size of our
simulation domain, is a dimensionless measure of the interfacial width; the
Schmidt number $Sc=\nu/D$. We assume that $\gamma$ and $\rho$ are independent 
of $\phi$; we keep the
diffusivity $D=\frac{\gamma\Lambda}{\xi^{2}}$ constant.  
We study the retraction dynamics of an antibubble by
varying the Atwood  number $A=(\rho_2-\rho_1)/(\rho_2+\rho_1)$ and the Bond
number $Bo=Ag\rho h_0^{2}/\sigma$. The Bond number is a convenient
dimensionless ratio of body forces (gravity), on the antibubble, and the
surface or interfacial tension $\sigma$.  We express times in multiples of
$\tau_g\equiv \sqrt{R_0/Ag}=1.16$.  

We solve the CHNS Eqs.\eqref{ns}-\eqref{ch} by using the pseudospectral method with
periodic boundary conditions; because of the cubic nonlinearity in the chemical
potential $\mu$, we use $N/2$-dealiasing~\cite{canuto2012spectral}.  We use the
exponential Adams-Bashforth method ETD2~\cite{cox2002exponential} for time
marching.  For our 2D DNSs, we use computers with Graphics Processing Units
(e.g., the NVIDIA K80), which we program in CUDA~\cite{cuda}. We use
conventional CPU-based computers for our 3D DNSs.  We have $N$ collocation
points in each direction, and our domain length is $L=2\pi$. Our efficient code
allows us to explore the CHNS parameter space and carry out very long
simulations that are essential for our studies.

%

For the VoF method we note the following points: It is a sharp-interface
method, so it has $Ch = 0$. The surface tension force for VoF is
$\mathbf{F}_{\sigma}\equiv \sigma \kappa \delta_s \mathbf{n}$, where $\kappa$
is the curvature, $\delta_s$ is the Dirac delta function on the interface, and
$\mathbf{n}$ is the normal to the interface.  We use Basilisk~\cite{basilisk,
popinet2018}, an open-source solver for carrying out both 2D and 3D
axisymmetric VoF DNSs.  Basilisk, which employs the Bell-Collela-Glaz advection
scheme~\cite{bell1989second} and the implicit viscosity solver~\cite{basilisk,
popinet2018}, is parallelized over conventional CPUs~\cite{basilisk,
popinet2018}. The VoF solver does not invoke the Boussinesq approximation; and
it can handle large density and viscosity contrasts. The breakup of an
interface is sensitive to the resolution in the VoF, so we use an initial
configuration in which the antibubble is already punctured at the bottom; and
we then investigate its spatiotemporal evolution. This initial condition is
similar to that used in experiments and in Ref.~\cite{sob2015theory}.  We
use the following boundary conditions in our VoF simulations: (a) In our 2D DNSs
we employ periodic boundary conditions in all directions; (b) in
the 3D axisymmetric case we use an axisymmetric boundary condition on the $z=0$
axis and the no-slip condition ${\bm u}=0$ at other boundaries. 

We list the parameters for some of our representative DNS runs in
Tables~\ref{table1a_chap7},~\ref{table1b_chap7},~\ref{table2_mod}, and
\ref{table3_pierce}. We also give Tables with the details of all our DNSs in
the Supplemental Material~\cite{supmat}.  As we show below, by utilising both
the CHNS framework and the VoF method, we can capture accurately the rupture
and retraction dynamics of an antibubble in an elegant and numerically
efficient way. 

\begin{table}
\resizebox{1.0\linewidth}{!}
{
\begin{tabular}{|l|l|l|l|l|l|l|l|l|l|l|l|}
\hline
& $N$ &$R_0/\xi$ & $R_1/\xi$ & $R_0/h_0$ & $h_0/\xi$ & $D$&$Ch$ &$Bo(\times 10^{-3})$ & $\nu$ &$Sc$ \\
\hline
\hline
{\tt {R1}} & $1024$ (2D) &$76.6$& $69.7$ & $7$ & $6.96$&$0.004$&$0.0028$&$36$ & $0.007$ &$1.75$\\
{\tt {R2}} & $1024$ (2D) &$76.6$& $69.7$ & $7$ & $6.96$&$0.004$&$0.0028$&$12$ & $0.007$ &$1.75$\\
{\tt {R3}} & $1024$ (2D) &$76.6$& $69.7$ & $7$ & $6.96$&$0.004$&$0.0028$&$8.9$ & $0.007$ &$1.75$\\
{\tt {R4}} & $1024$ (2D) &$76.6$& $69.7$ & $7$ & $6.96$&$0.004$&$0.0028$&$5.0$ & $0.007$ &$1.75$\\
{\tt {R5}} & $1024$ (2D) &$76.6$& $69.7$ & $7$ & $6.96$&$0.004$&$0.0028$&$4.0$ & $0.007$ &$1.75$\\
{\tt {R6}} & $1024$ (2D) &$76.6$& $69.7$ & $7$ & $6.96$&$0.004$&$0.0028$&$4.5$ & $0.007$ &$1.75$\\
{\tt {R7}} & $1024$ (2D) &$76.6$& $69.7$ & $7$ & $6.96$&$0.004$&$0.0028$&$3.3$ & $0.007$ &$1.75$\\
{\tt {R8}} & $1024$ (2D) &$76.6$& $69.7$ & $7$ & $6.96$&$0.004$&$0.0028$&$2.7$ & $0.007$ &$1.75$\\
{\tt {R9}} & $1024$ (2D) &$76.6$& $69.7$ & $7$ & $6.96$&$0.004$&$0.0028$&$2.4$ & $0.007$ &$1.75$\\
{\tt {R10}} & $1024$ (2D) &$76.6$& $69.7$ & $7$ & $6.96$&$0.004$&$0.0028$&$2.1$ &$0.007$ & $1.75$\\
{\tt {R11}} & $1024$ (2D) &$76.6$& $69.7$ & $7$ & $6.96$&$0.004$&$0.0028$&$1.9$ & $0.007$ &$1.75$\\
{\tt {R12}} & $1024$ (2D) &$76.6$& $69.7$ & $7$ & $6.96$&$0.004$&$0.0028$&$1.7$ & $0.007$ &$1.75$\\
{\tt {R13}} & $1024$ (2D) &$76.6$& $69.7$ & $7$ & $6.96$&$0.004$&$0.0028$&$1.5$ & $0.007$ &$1.75$\\
{\tt {R14}} & $1024$ (2D) &$76.6$& $69.7$ & $7$ & $6.96$&$0.004$&$0.0028$&$1.4$ & $0.007$ &$1.75$\\
{\tt {R15}} & $1024$ (2D) &$76.6$& $69.7$ & $7$ & $6.96$&$0.004$&$0.0028$&$1.3$ & $0.007$ &$1.75$\\
{\tt {R16}} & $1024$ (2D) &$76.6$& $69.7$ & $7$ & $6.96$&$0.004$&$0.0028$&$1.1$ & $0.007$ &$1.75$\\
{\tt {R17}} & $1024$ (2D) &$76.6$& $69.7$ & $7$ & $6.96$&$0.004$&$0.0028$&$0.8$ & $0.007$ &$1.75$\\
{\tt {R18}} & $1024$ (2D) &$76.6$& $69.7$ & $7$ & $6.96$&$0.004$&$0.0028$&$0.73$ & $0.007$ &$1.75$\\
{\tt {R19}} & $1024$ (2D) &$76.6$& $69.7$ & $7$ & $6.96$&$0.004$&$0.0028$&$0.56$ & $0.007$ &$1.75$\\
{\tt {R20}} & $1024$ (2D) &$76.6$& $69.7$ & $7$ & $6.96$&$0.004$&$0.0028$&$0.45$ & $0.007$ &$1.75$\\
\hline
{\tt {R21}} & $1024$ (2D) &$76.6$& $69.7$ & $8$ & $6.96$&$0.004$&$0.0028$&$12$ & $0.007$ &$1.75$\\
{\tt {R22}} & $1024$ (2D) &$76.6$& $69.7$ & $8$ & $6.96$&$0.004$&$0.0028$&$5.0$ & $0.007$ &$1.75$\\
{\tt {R23}} & $1024$ (2D) &$76.6$& $69.7$ & $8$ & $6.96$&$0.004$&$0.0028$&$4.0$ & $0.007$ &$1.75$\\
{\tt {R24}} & $1024$ (2D) &$76.6$& $69.7$ & $8$ & $6.96$&$0.004$&$0.0028$&$3.3$ & $0.007$ &$1.75$\\
{\tt {R25}} & $1024$ (2D) &$76.6$& $69.7$ & $8$ & $6.96$&$0.004$&$0.0028$&$3.1$ & $0.007$ &$1.75$\\
{\tt {R26}} & $1024$ (2D) &$76.6$& $69.7$ & $8$ & $6.96$&$0.004$&$0.0028$&$2.7$ & $0.007$ &$1.75$\\
{\tt {R27}} & $1024$ (2D) &$76.6$& $69.7$ & $8$ & $6.96$&$0.004$&$0.0028$&$2.4$ & $0.007$ &$1.75$\\
{\tt {R28}} & $1024$ (2D) &$76.6$& $69.7$ & $8$ & $6.96$&$0.004$&$0.0028$&$1.9$ & $0.007$ &$1.75$\\
{\tt {R29}} & $1024$ (2D) &$76.6$& $69.7$ & $8$ & $6.96$&$0.004$&$0.0028$&$1.7$ & $0.007$ &$1.75$\\
{\tt {R30}} & $1024$ (2D) &$76.6$& $69.7$ & $8$ & $6.96$&$0.004$&$0.0028$&$1.5$ & $0.007$ &$1.75$\\
{\tt {R31}} & $1024$ (2D) &$76.6$& $69.7$ & $8$ & $6.96$&$0.004$&$0.0028$&$1.3$ & $0.007$ &$1.75$\\
{\tt {R32}} & $1024$ (2D) &$76.6$& $69.7$ & $8$ & $6.96$&$0.004$&$0.0028$&$1.1$ & $0.007$ &$1.75$\\
{\tt {R33}} & $1024$ (2D) &$76.6$& $69.7$ & $8$ & $6.96$&$0.004$&$0.0028$&$0.8$ & $0.007$ &$1.75$\\
{\tt {R34}} & $1024$ (2D) &$76.6$& $69.7$ & $8$ & $6.96$&$0.004$&$0.0028$&$0.73$ & $0.007$ &$1.75$\\
{\tt {R35}} & $1024$ (2D) &$76.6$& $69.7$ & $8$ & $6.96$&$0.004$&$0.0028$&$0.45$ & $0.007$ &$1.75$\\
\hline
\end{tabular}
}
\caption{The parameters $R_0,R_1,h_0,\xi$, $\sigma$,
$\nu,D,CH,Sc$ for our CHNS DNS runs
{\tt {R1-R35}}. Initially, $R_0$ and $R_1$ are, respectively,
the outer and inner radii of the antibubble; $h_0 \equiv (R_0-R_1)$;  
$\sigma$ is the surface tension; $\nu$ is the kinematic
viscosity. In all the runs, the Atwood number times the acceleration due to gravity 
$Ag=0.99$; the number of collocation points is $N^2$;
the diffusivity is $D$; the Cahn number $Ch=\xi/L$, $\xi$ being the interface width.  
The typical values of $R_0/h_0$ used in experiments are about 
$38$~\cite{zou2013collapse}. For all our 2D runs, the friction coefficient 
$\alpha=0.001$.}
\label{table1a_chap7} 
\end{table}

\begin{table}
\resizebox{1.0\linewidth}{!}
{
\begin{tabular}{|l|l|l|l|l|l|l|l|l|l|l|l|}
\hline
& $N$ &$R_0/\xi$ & $R_1/\xi$ & $R_0/h_0$ & $h_0/\xi$ & $D$&$Ch$ &$Bo(\times 10^{-3})$ & $\nu$ &$Sc$ \\
\hline
\hline
{\tt {R36}} & $1024$ (2D) &$76.6$& $69.7$ & $9$ & $6.96$&$0.004$&$0.0028$&$12$ & $0.007$ &$1.75$\\
{\tt {R37}} & $1024$ (2D) &$76.6$& $69.7$ & $9$ & $6.96$&$0.004$&$0.0028$&$5.0$ & $0.007$ &$1.75$\\
{\tt {R38}} & $1024$ (2D) &$76.6$& $69.7$ & $9$ & $6.96$&$0.004$&$0.0028$&$4.0$ & $0.007$ &$1.75$\\
{\tt {R39}} & $1024$ (2D) &$76.6$& $69.7$ & $9$ & $6.96$&$0.004$&$0.0028$&$3.3$ & $0.007$ &$1.75$\\
{\tt {R40}} & $1024$ (2D) &$76.6$& $69.7$ & $9$ & $6.96$&$0.004$&$0.0028$&$3.1$ & $0.007$ &$1.75$\\
{\tt {R41}} & $1024$ (2D) &$76.6$& $69.7$ & $9$ & $6.96$&$0.004$&$0.0028$&$2.7$ & $0.007$ &$1.75$\\
{\tt {R42}} & $1024$ (2D) &$76.6$& $69.7$ & $9$ & $6.96$&$0.004$&$0.0028$&$2.4$ & $0.007$ &$1.75$\\
{\tt {R43}} & $1024$ (2D) &$76.6$& $69.7$ & $9$ & $6.96$&$0.004$&$0.0028$&$1.9$ & $0.007$ &$1.75$\\
{\tt {R44}} & $1024$ (2D) &$76.6$& $69.7$ & $9$ & $6.96$&$0.004$&$0.0028$&$1.7$ & $0.007$ &$1.75$\\
{\tt {R45}} & $1024$ (2D) &$76.6$& $69.7$ & $9$ & $6.96$&$0.004$&$0.0028$&$1.5$ & $0.007$ &$1.75$\\
{\tt {R46}} & $1024$ (2D) &$76.6$& $69.7$ & $9$ & $6.96$&$0.004$&$0.0028$&$1.3$ & $0.007$ &$1.75$\\
{\tt {R47}} & $1024$ (2D) &$76.6$& $69.7$ & $9$ & $6.96$&$0.004$&$0.0028$&$1.1$ & $0.007$ &$1.75$\\
{\tt {R48}} & $1024$ (2D) &$76.6$& $69.7$ & $9$ & $6.96$&$0.004$&$0.0028$&$0.8$ & $0.007$ &$1.75$\\
{\tt {R49}} & $1024$ (2D) &$76.6$& $69.7$ & $9$ & $6.96$&$0.004$&$0.0028$&$0.73$ & $0.007$ &$1.75$\\
{\tt {R50}} & $1024$ (2D) &$76.6$& $69.7$ & $9$ & $6.96$&$0.004$&$0.0028$&$0.45$ & $0.007$ &$1.75$\\
\hline
{\tt {R51}} & $1024$ (2D) &$76.6$& $69.7$ & $11$ & $6.96$&$0.004$&$0.0028$&$0.01$ & $0.007$ &$1.75$\\
{\tt {R52}} & $1024$ (2D) &$76.6$& $69.7$ & $11$ & $6.96$&$0.004$&$0.0028$&$0.01$ & $0.01$ &$2.6$\\
{\tt {R53}} & $1024$ (2D) &$76.6$& $69.7$ & $11$ & $6.96$&$0.004$&$0.0028$&$0.01$ & $0.01$ &$3.9$\\
{\tt {R54}} & $1024$ (2D) &$76.6$& $69.7$ & $11$ & $6.96$&$0.004$&$0.0028$&$0.01$ & $0.02$ &$5.0$\\
{\tt {R55}} & $1024$ (2D) &$76.6$& $69.7$ & $11$ & $6.96$&$0.004$&$0.0028$&$0.01$ & $0.02$ &$6.8$\\
{\tt {R56}} & $1024$ (2D) &$76.6$& $69.7$ & $11$ & $6.96$&$0.004$&$0.0028$&$0.01$ & $0.035$ &$8.75$\\
{\tt {R57}} & $1024$ (2D) &$76.6$& $69.7$ & $11$ & $6.96$&$0.004$&$0.0028$&$0.01$ & $0.035$ &$10.0$\\
{\tt {R58}} & $1024$ (2D) &$76.6$& $69.7$ & $11$ & $6.96$&$0.004$&$0.0028$&$0.01$ & $0.05$ &$12.5$\\
{\tt {R60}} & $1024$ (2D) &$76.6$& $69.7$ & $11$ & $6.96$&$0.004$&$0.0028$&$0.01$ & $0.05$ &$14.0$\\
{\tt {R61}} & $1024$ (2D) &$76.6$& $69.7$ & $11$ & $6.96$&$0.004$&$0.0028$&$0.01$ & $0.05$ &$15.75$\\
{\tt {R62}} & $1024$ (2D) &$76.6$& $69.7$ & $11$ & $6.96$&$0.004$&$0.0028$&$0.01$ & $0.05$ &$17.5$\\
{\tt {R63}} & $1024$ (2D) &$76.6$& $69.7$ & $11$ & $6.96$&$0.004$&$0.0028$&$0.01$ & $0.08$ &$20$\\
{\tt {R64}} & $1024$ (2D) &$76.6$& $69.7$ & $11$ & $6.96$&$0.004$&$0.0028$&$0.01$ & $0.12$ &$30$\\
{\tt {R65}} & $1024$ (2D) &$76.6$& $69.7$ & $11$ & $6.96$&$0.004$&$0.0028$&$0.01$ & $0.18$ &$45$\\
{\tt {R66}} & $1024$ (2D) &$76.6$& $69.7$ & $11$ & $6.96$&$0.004$&$0.0028$&$0.01$ & $0.18$ &$58$\\
{\tt {R67}} & $1024$ (2D) &$76.6$& $69.7$ & $11$ & $6.96$&$0.004$&$0.0028$&$0.01$ & $0.27$ &$67.5$\\
{\tt {R68}} & $1024$ (2D) &$76.6$& $69.7$ & $11$ & $6.96$&$0.004$&$0.0028$&$0.01$ & $0.27$ &$80.0$\\
{\tt {R69}} & $1024$ (2D) &$76.6$& $69.7$ & $11$ & $6.96$&$0.004$&$0.0028$&$0.01$ & $0.4$ &$100$\\
{\tt {R70}} & $1024$ (2D) &$76.6$& $69.7$ & $11$ & $6.96$&$0.004$&$0.0028$&$0.01$ & $0.6$ &$151.5$\\
{\tt {R71}} & $1024$ (2D) &$76.6$& $69.7$ & $11$ & $6.96$&$0.004$&$0.0028$&$0.01$ & $0.96$ &$241.7$\\
{\tt {R72}} & $1024$ (2D) &$76.6$& $69.7$ & $11$ & $6.96$&$0.004$&$0.0028$&$0.1$ & $0.96$ &$241.7$\\
\hline
\end{tabular}
}
\caption{The parameters $R_0,R_1,h_0,\xi$, $\sigma$,
$\nu,D,CH,Sc$ for our CHNS DNS runs
{\tt {R36-R72}}. Initially, $R_0$ and $R_1$ are, respectively,
the outer and inner radii of the antibubble; $h_0 \equiv (R_0-R_1)$;  
$\sigma$ is the surface tension; $\nu$ is the kinematic
viscosity. In all the runs, the Atwood number times the acceleration due to gravity 
$Ag=0.99$, the number of collocation points is $N^2$;
the diffusivity is $D$; the Cahn number $Ch=\xi/L$, $\xi$ being the interface width.  
The typical values of $R_0/h_0$ used in experiments are about 
$38$~\cite{zou2013collapse}. For all our 2D runs, the friction coefficient $\alpha=0.001$.}
\label{table1b_chap7} 
\end{table}

\begin{table}
\resizebox{1.0\linewidth}{!}
{
\begin{tabular}{|l|l|l|l|l|l|l|l|l|l|l|l|}
\hline
& $N$ &$R_0/\xi$ & $R_1/\xi$ & $R_0/h_0$ & $h_0/\xi$ & $D$&$Ch$ &$Bo(\times 10^{-3})$ & $\nu$ &$Sc$ \\
\hline
\hline
{\tt{Q1}} & $256$ (3D)&$24.0$ & $12.0$ & $1.2$& $12.0$ & $0.006$&$0.013$ & $1.3$&$0.0016$ &$0.27$\\
{\tt{Q2}} & $256$ (3D)&$24.0$ & $12.0$ & $1.2$& $12.0$ & $0.006$&$0.013$ & $0.63$&$0.0016$ &$0.27$\\
{\tt{Q3}} & $256$ (3D)&$24.0$ & $12.0$ & $1.2$& $12.0$ & $0.006$&$0.013$ & $0.4$&$0.0016$ &$0.27$\\
{\tt{Q4}} & $256$ (3D)&$24.0$ & $12.0$ & $1.2$& $12.0$ & $0.006$&$0.013$ & $0.3$&$0.0016$ &$0.27$\\
{\tt{Q5}} & $256$ (3D)&$24.0$ & $12.0$ & $1.2$& $12.0$ & $0.006$&$0.013$ & $0.25$&$0.0016$ &$0.27$\\
{\tt{Q6}} & $256$ (3D)&$24.0$ & $12.0$ & $1.2$& $12.0$ & $0.006$&$0.013$ & $0.21$&$0.0016$ &$0.27$\\
{\tt{Q7}} & $512$ (3D)& $14$ & $10.5$ & $4.0$&$3.5$& $0.006$&$0.011$&$2.7$& $0.0016$&$0.27$\\
\hline
\end{tabular}
}
\caption{The parameters $R_0,R_1,h_0,\xi$, $\sigma$,
$\nu,D,CH,Sc$ for our CHNS DNS runs {\tt{Q1-Q7}}. Initially, $R_0$ and $R_1$ are, respectively,
the outer and inner radii of the antibubble; $h_0 \equiv (R_0-R_1)$;  
$\sigma$ is the surface tension; $\nu$ is the kinematic
viscosity. In all the runs, the Atwood number times the acceleration due to gravity 
$Ag=0.5$, the number of collocation points is for {\tt {Q1-Q6}} is $N^3=256^3$ and 
for {\tt {Q7}} is $N^3=512^3$: the diffusivity is $D$; the Cahn number $Ch=\xi/L$, 
with $\xi$ being the interface width. The typical values of $R_0/h_0$ used in 
experiments are about $38$~\cite{zou2013collapse}.}
\label{table2_mod} 
\end{table}

%
%

\begin{table}
\resizebox{1.0\linewidth}{!}
{
\begin{tabular}{|l|l|l|l|l|l|l|l|l|}
\hline
& $R_0$ & $R_1$& $h_0$&$A$ & $g$& $\nu$ &$Bo(\times 10^{-3})$\\
\hline
\hline
{\tt {PR1}} & $0.859$& $0.739$  & $0.12$& $0.9$&$1.$& $0.007$ &$0.9$\\
{\tt {PR2}} & $0.859$& $0.739$  & $0.12$& $0.9$&$1.$& $0.007$ &$1.17$\\
{\tt {PR3}} & $0.859$& $0.739$  & $0.12$& $0.9$&$1.$& $0.007$ &$1.82$\\
{\tt {PR4}} & $0.859$& $0.739$  & $0.12$& $0.9$&$1.$& $0.007$ &$4.1$\\
{\tt {PR5}} & $0.859$& $0.739$  & $0.12$& $0.9$&$1.$& $0.007$ &$8.2$\\
{\tt {PR6}} & $0.859$& $0.739$  & $0.12$& $0.9$&$1.$& $0.007$ &$16.4$\\
{\tt {APR1}} & $0.859$& $0.739$  & $0.12$& $0.9$&$1.$& $0.007$ &$0.75$\\
{\tt {APR2}} & $0.859$& $0.739$  & $0.12$& $0.9$&$1.$& $0.007$ &$0.9$\\
{\tt {APR3}} & $0.859$& $0.739$  & $0.12$& $0.9$&$1.$& $0.007$ &$1.17$\\
{\tt {APR4}} & $0.859$& $0.739$  & $0.12$& $0.9$&$1.$& $0.007$ &$1.82$\\
{\tt {APR5}} & $0.859$& $0.739$  & $0.12$& $0.9$&$1.$& $0.007$ &$2.05$\\
{\tt {APR6}} & $0.859$& $0.739$  & $0.12$& $0.9$&$1.$& $0.007$ &$4.1$\\
{\tt {APR7}} & $0.859$& $0.739$  & $0.12$& $0.9$&$1.$& $0.007$ &$8.2$\\
{\tt {APR8}} & $0.859$& $0.739$  & $0.12$& $0.9$&$1.$& $0.007$ &$16.4$\\
{\tt {APR9}} & $0.859$& $0.739$  & $0.12$& $0.01$&$100.$&$0.007$ &$0.9$\\
{\tt {APR10}} & $0.859$& $0.739$  & $0.12$& $0.01$&$100.$&$0.007$ &$1.82$\\
{\tt {APR11}} & $0.859$& $0.739$  & $0.12$& $0.01$&$100.$&$0.007$ &$2.05$\\
{\tt {APR12}} & $0.859$& $0.739$  & $0.12$& $0.01$&$100.$&$0.007$ &$4.1$\\
{\tt {APR13}} & $0.859$& $0.739$  & $0.12$& $0.01$&$100.$&$0.007$ &$8.2$\\
{\tt {APR14}} & $0.859$& $0.739$  & $0.12$& $0.01$&$100.$&$0.007$ &$16.4$\\
\hline
\end{tabular}
}
\caption{\label{table3_pierce} The parameters  $R_0, R_1, h_0$, $A$, $g$, $\nu$
and $Bo$ for our VoF runs: 2D(${\tt {PR1}}-{\tt {PR6}}$); and 3D axisymmetric
(${\tt {APR1}}-{\tt {APR14}}$). For these runs $\rho_2 =1$. For the 2D and 
axisymmetric 3D runs we use a square box of area $L^2=16$ and discretize it with 
$N^2=2048^2$ collocation points. For our 2D VoF DNSs we employ
periodic boundary conditions in all directions; in 3D we use an  axisymmetric 
boundary condition on the $z=0$ axis and the no-slip boundary condition ${\bm u}=0$  
at other boundaries.}
\end{table} 

\section{Results}
\label{sec:Results}

We now present the results of our DNSs in the following subsections: In
subsection~\ref{subsec:SpTem2D} we illustrate the spatiotemporal evolution of
2D antibubbles and, in subsection~\ref{subsec:Spectra2D}, we give the temporal
evolution of Fourier spectra of $\phi$ and the velocity and vorticity fields
that are associated with these antibubbles. In subsection~\ref{subsec:Rim} we
explore the dependence on $\sigma$ of the velocity of the retracting rim of a
ruptured 2D antibubble; and, in subsection~\ref{subsec:EnTime}, we examine the
signatures of this rupture in the time dependence of the energy. Furthermore,
we elucidate the dependence of the antibubble-rupture time on the size of the
antibubble and the surface tension (subsection~\ref{subsec:Size}) and on the
kinematic viscosity (subsection~\ref{subsec:Viscosity}).  We then present
illustrative results on the spatiotemporal evolution of 3D antibubbles in
subsection~\ref{subsec:SpTem3D}  

\REM{
\begin{figure*}
\includegraphics[width=.325\linewidth]{images_phase_small/t00000001}
\put(-60,40){\bf \scriptsize \textcolor{white}{(a)}}
\includegraphics[width=.325\linewidth]{images_phase_small/t00000061}
\put(-60,40){\bf \scriptsize \textcolor{white}{(b)}}
\includegraphics[width=.325\linewidth]{images_phase_small/t00000220}
\put(-60,40){\bf \scriptsize \textcolor{white}{(c)}}

\includegraphics[width=.325\linewidth]{images_vort_2d/t00000001}
\put(-60,40){\bf \scriptsize(d)}
\includegraphics[width=.325\linewidth]{images_vort_2d/t00000061}
\put(-60,40){\bf \scriptsize(e)}
\includegraphics[width=.325\linewidth]{images_vort_2d/t00000220}
\put(-60,40){\bf \scriptsize(f)}

\caption{Pseudocolor plots of the CHNS field $\phi$ with antibubble radius
$R_0/h_0=7$, $\nu=0.007$ and $\sigma=16.6$ (i.e, $Bo=1.84 \times 10^{-4}$, from
our 2D DNS run {\tt{R2}}) at (a) $t=1.07\times 0.05\tau_g$, (b)
$t=65.27\times 0.05\tau_g$, (c)
$t=235.4\times 0.05\tau_g$; pseudocolor plots of the corresponding $\omega$ field at (d)
$t=1.07\times 0.05\tau_g$, (e) $t=65.27\times 0.05\tau_g$, and (f) $t=235.4\times 0.05\tau_g$.  We show all the values of time
$t$ in multiples of $\tau_g$. These plots show that, initially, the antibubble
rises under gravity, breaks, (a,d), then it retracts (b,e),  becomes a circular 
droplet, and rises again (e,f). We show all the values of the time $t$
in multiples of $0.05\tau_g$.
} \label{2dinitial}
\end{figure*}
}

\begin{figure*}
\includegraphics[width=0.85\linewidth]{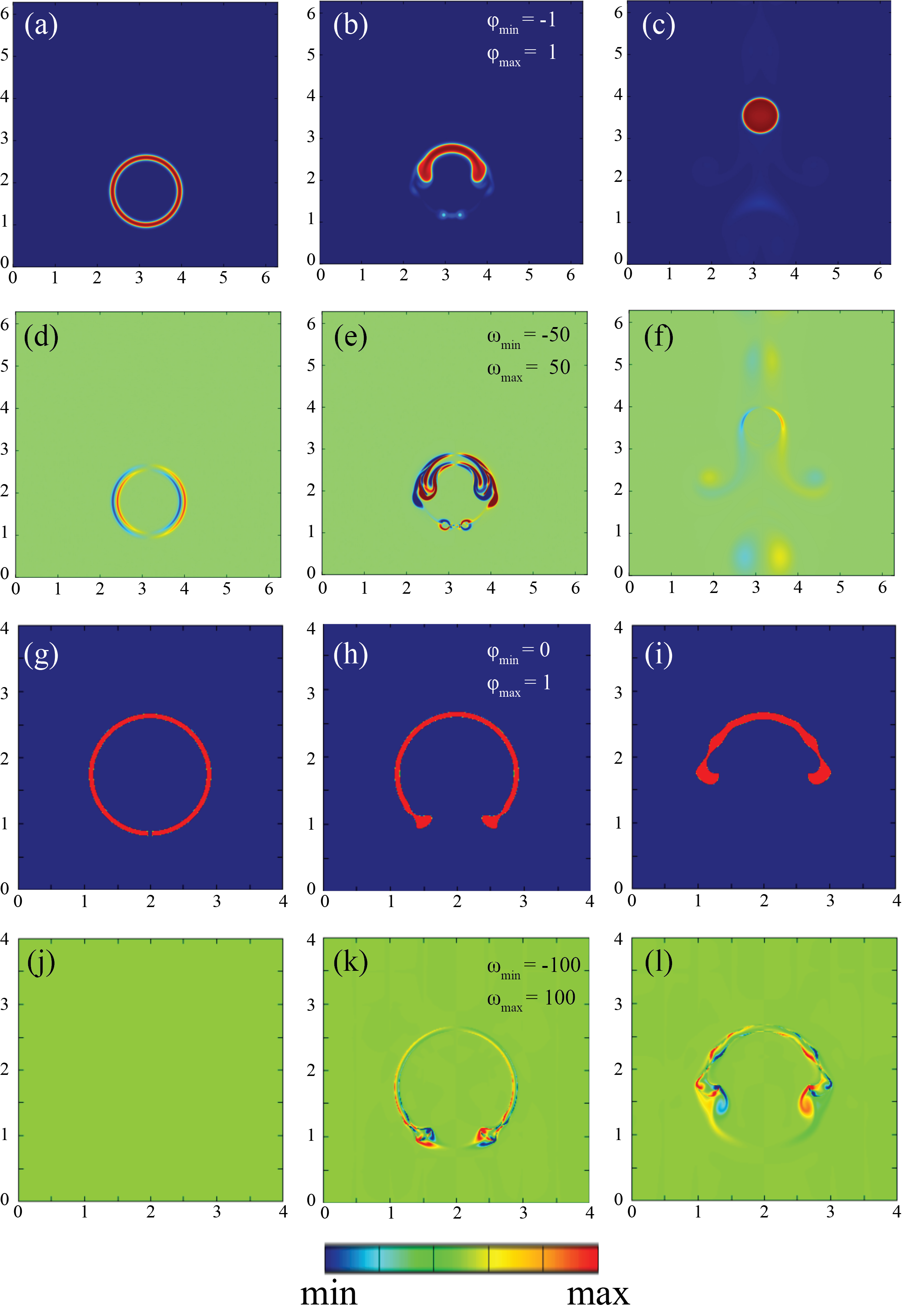}
\caption{\label{cont_plots_pierce}  Pseudocolor plots of the CHNS $\phi$ field 
with antibubble radius $R_0/h_0=7$ at time (a) $t=0.05\tau_g$, (b) $t=3.26\tau_g$, (c) $t=11.77\tau_g$ for $\nu=0.007$ and $\sigma=16.6$ (i.e, $Bo=0.001$, from
our 2D DNS run {\tt{R16}}); pseudocolor plots of the corresponding $\omega$ field (d)--(f). These plots show that, initially, the antibubble
rises under gravity, breaks, (a,d), then it retracts (b,e),  becomes a circular 
droplet and rises again (e,f). Pseudocolor plots of the volume fraction 
for the 2D VoF run {\tt{PR5}} at time (g) $t=0$, (h) $t=0.1\tau_g$, (i) $t=0.3\tau_g$; 
(j), (k), and (l) pseudocolor plots of the vorticity for the 2D VoF  
{\tt{PR5}} run at the same times as in (g), (h), and (i), respectively;
in these 2D VoF runs, $\sigma = 1.68$; these plots show that 
the antibubble retracts and forms a rim. Time is measured in units of $\tau_g$. }
\end{figure*}


\subsection{Spatiotemporal evolution: 2D} 
\label{subsec:SpTem2D} 

%
%

%
%

In Fig.~\ref{cont_plots_pierce} we display, via pseudocolor plots of $\phi$ and
$\omega$, the spatiotemporal evolution of a 2D antibubble rising under gravity
in our CHNS DNS; blue and red indicate heavy and light fluids, respectively.  
Figure~\ref{cont_plots_pierce}(a) shows the
pseudocolor plot of $\phi$ for the antibubble at a time $\simeq 1.07\times
0.05\tau_g$. Initially, the antibubble rises because of gravity, breaks
(Fig.~\ref{cont_plots_pierce}(b)), then displays arms that retract, forms a
disc-shaped droplet, which rises under gravity
(Fig.~\ref{cont_plots_pierce}(c)), because the antibubble-shell (red) fluid is
lighter than the background (blue) fluid.
Figures~\ref{cont_plots_pierce}(d)-(f) show pseudocolor plots of the vorticity
field at the same times as their $\phi$ counterparts in
Figs.~\ref{cont_plots_pierce}(a)-(c). The spatiotemporal evolution of such a 2D
antibubble is shown in the Videos~\ref{sup_video1} and~\ref{sup_video2} in the Supplemental Material~\cite{supmat}.
Note that, just after forming the disc-type shape, the antibubble goes down in
a direction opposite to that dictated by gravity
(Fig.~\ref{cont_plots_pierce}(e)), because the surface tension is high, so,
during the retraction of the arms, when interfacial energy is released, the
retracting droplet can be pushed in this opposite direction. Eventually this
backward thrust is damped by viscosity and, finally, the (light) droplet rises
under gravity.  We begin with $\omega=0$, everywhere. We see that, because
of the $\nabla \times \phi\nabla\mu$ term on the right-hand side (RHS) of
Eq.\eqref{ns_2d}, a vorticity field is generated at the outer and inner surfaces
of the antibubble, i.e., wherever $\phi$ changes in space.  The vortical
regions that are generated initially have a size that is comparable to the
interface width.  These vortices then grow with time, until they are damped by
viscosity. Similarly, the gravity term is also significant wherever $\phi$
changes sign. The competition between the viscous, surface-tension, and gravity
terms in the CHNS equations determines the breakup time of an antibubble, as we
show below.

In Figs.~\ref{cont_plots_pierce}(g)-(l) we use pseudocolor plots to illustrate
antibubble retraction in our VoF DNS in 2D (run {\tt{PR1}}).  Both our CHNS and
VoF DNSs show that, as it ruptures, an antibubble forms a rim, which then
retracts.

\begin{figure*}
\includegraphics[width=0.85\linewidth]{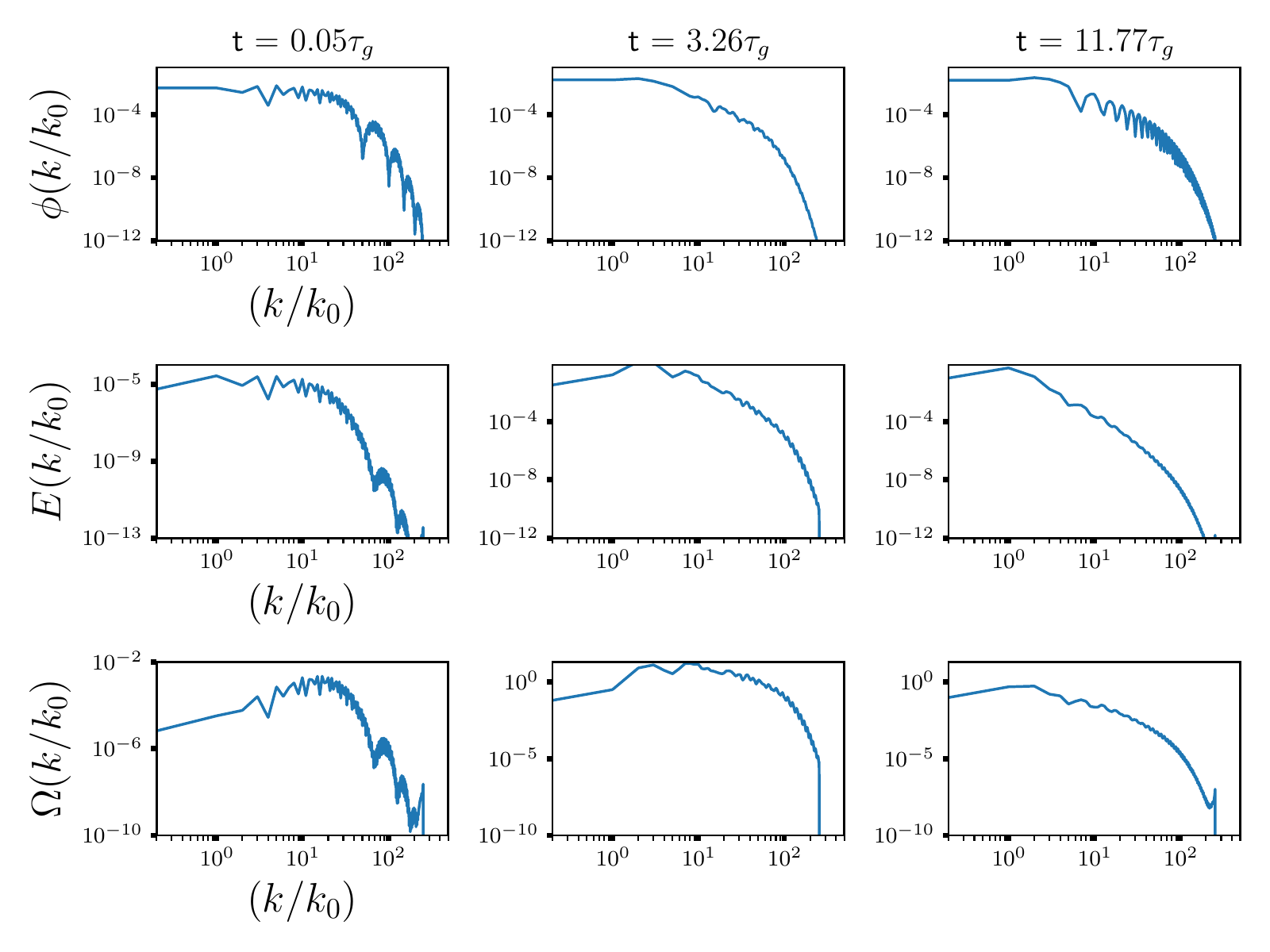}
\put(-300,290){\bf \scriptsize {(a)}}
\put(-160,290){\bf \scriptsize {(b)}}
\put(-25,290){\bf \scriptsize {(c)}}
\put(-300,190){\bf \scriptsize {(d)}}
\put(-160,190){\bf \scriptsize {(e)}}
\put(-25,190){\bf \scriptsize {(f)}}
\put(-300,90){\bf \scriptsize {(g)}}
\put(-160,90){\bf \scriptsize {(h)}}
\put(-25,90){\bf \scriptsize {(i)}}
	\caption{\label{spectra_anti} Log--log (base 10) plots of the spectra 
	(Eq.~\ref{spec}) (a)--(c) $S(k,t)$, (d)--(f) $E(k,t)$, and  (g)--(i) 
	$\Omega(k,t)$ versus $k$, with $k_0=1$ and time in units of $\tau_g$. }
\end{figure*}

\subsection{Fourier spectra}
\label{subsec:Spectra2D}

Fourier-space spectra give us a complementary view of the spatiotemporal
evolution of the rupture of an antibubble (cf., the physical-space pseudocolor
plots in Figs.~\ref{cont_plots_pierce} (a)-(l)). In Fig.~\ref{spectra_anti} we show,
for our 2D CHNS DNSs, log-log (base 10) plots of the spectra for the field $\phi$,
the energy spectrum $E(k,t)$, and the enstrophy spectrum $\Omega(k,t)$, at
different times during this evolution. These spectra are defined as follows:
\begin{eqnarray}
	S(k,t) &\equiv& \sum \limits_{k-\frac{1}{2}\leq k^{\prime}\leq k+\frac{1}{2}}\langle \hat{\phi}({\bm k}^{\prime},t) \hat{\phi}(-{\bm k}^{\prime},t) \rangle ; \nonumber \\
	E(k,t) &\equiv& \df{1}{2}\sum \limits_{k-\frac{1}{2}\leq k^{\prime}\leq k+\frac{1}{2}}\langle \hat{\bm{u}}({\bm k}^{\prime},t)\hat{\bm{u}}(-{\bm k}^{\prime},t) \rangle ; \nonumber \\
	\Omega(k,t) &\equiv&  \df{1}{2}\sum \limits_{k-\frac{1}{2}\leq k^{\prime}\leq k+\frac{1}{2}}\langle \hat{{\omega}}({\bm k}^{\prime},t)\hat{{\omega}}(-{\bm k}^{\prime},t) \rangle ;
\label{spec}
\end{eqnarray}
here, the circumflex denotes a spatial Fourier transform and $k$ and
$k^{\prime}$ are, respectively, the magnitudes of the wave vectors ${\bm k}$
and ${\bm k}^{\prime}$. At early times (Figs.~\ref{spectra_anti} (a), (d), and
(g)), these spectra, especially $S(k,t)$, show oscillations with small and
large periods, which are, respectively, related inversely to the outer radius
of the antibubble and the thickness of its shell.  As the antibubble ruptures
(Figs.~\ref{spectra_anti} (b), (e), and (h)), it loses its circular shape, and,
in turn, the spectra lose their oscillations.  Finally, the antibubble is
replaced by a single droplet (Fig.~\ref{cont_plots_pierce}(c)) of the light
fluid, which rises under gravity; at this stage, these spectra, especially
$S(k,t)$, show oscillations with a small period, which is related inversely to
the radius of this rising droplet. These Fourier-space spectra show that the
breakup of an antibubble leads to turbulence, insofar as $E(k,t)$, $S(k,t)$,
and $\Omega(k,t)$ extend significantly over several orders of magnitude of $k$.
The Taylor-microscale Reynolds number $Re_\lambda$ is $0.54$, $128.5$, and
$163.4$ in Figs.~\ref{spectra_anti} (a), (b), and (c), respectively; here,
$Re_\lambda = u_{rms} \lambda /\nu$, where the root-mean-square velocity
$u_{rms}(t) = 2\sum \limits_{k} E(k,t)$ and the Taylor microscale $\lambda
\equiv [\sum k^2 E(k,t)/\sum E(k,t)]^{-1/2}$.

\begin{center}
\begin{figure}
\includegraphics[width=.8\linewidth]{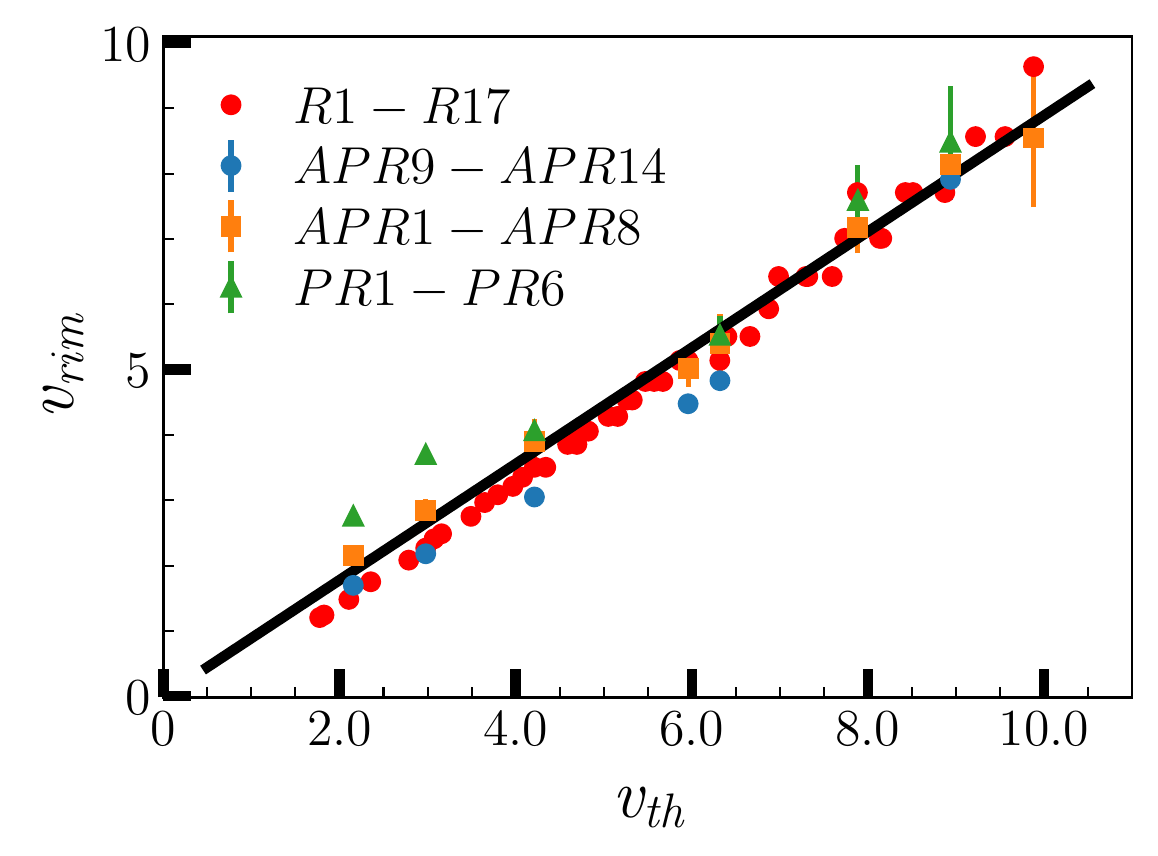}
\put(-80,70)

\caption{\label{sigma_tlife} Plot of the retraction velocity $v_{rim}$ versus the 
theoretical estimate $v_{th}=\sqrt{\sigma/\rho_2 a}$ for our 2D CHNS runs {\tt{R1-R17}},
2D VoF runs {\tt{PR1-PR6}}, and 3D axisymmetric VoF runs {\tt{APR1-APR14}};
red circles show data from our CHNS runs; green,
yellow, and blue markers indicate data from our VoF runs. The black line represents the 
theoretical line. Here the surface tension $\sigma=\displaystyle\frac{Ag\rho h_0^2}{Bo}$, 
where $Bo$ is the Bond number.
}
\end{figure}
\end{center}

\begin{figure*}
\includegraphics[width=.9\linewidth]{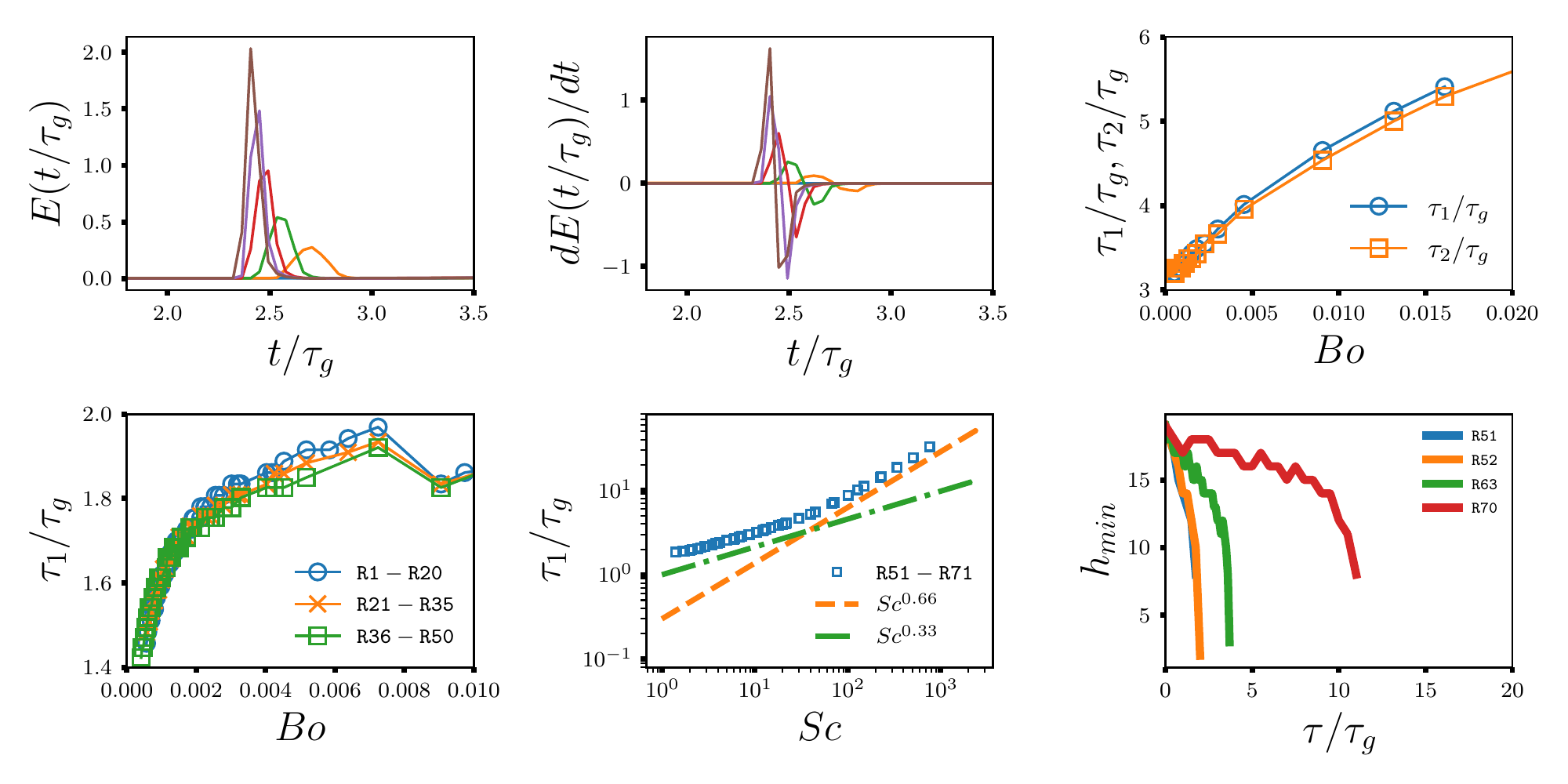}
\put(-420,210){\bf (a)}
\put(-265,210){\bf (b)}
\put(-115,210){\bf (c)}
\put(-420,100){\bf (d)}
\put(-265,100){\bf (e)}
\put(-115,103){\bf \scriptsize(f)}
\caption{Plots of (a) the total kinetic energy $E(t)$ and (b) the time derivative 
$dE(t)/dt$ versus $t/\tau_g$ for the 2D CHNS runs {\tt{R16-R20}};
(c) plots versus $Bo$ of the antibubble lifetimes $\tau_1$ and $\tau_2$ calculated, 
respectively, from the energy time series (blue line with circles)
and the minimum antibubble thickness (orange line with squares), for the 
2D CHNS runs {\tt{R1-R16}}.
(d) Plots versus $Bo$ (for $Sc=1.75$), of $\tau_1/\tau_g$ 
for $R_0/h_0=7$ (DNS run {\tt{R1-R20}}, blue
line with circles), $R_0/h_0=8$ (DNS run {\tt{R21-R35}}, orange line with crosses), 
$R_0/h_0=9$ (DNS run {\tt{R36-R50}}, green line with squares); 
(e) log-log plot of $\tau_1/\tau_g$ versus $Sc$, over three orders of magnitude
and at a fixed value of $Bo$ (runs {\tt R51-R71}); (f) plots of $h_{min}$ versus $t/\tau_g$, at $Bo=0.01$ and $A=0.01$ and $g=99$, for $Sc=1.75$ ({\tt R51}, blue line), $Sc=1.75$ ({\tt R52}, orange line)
	$Sc=20$ ({\tt R63}, green line), and $Sc=151.5$ ({\tt R70}, red line).}

\label{sigma_tlife2}
\end{figure*}

\REM
{
\begin{figure*}
\includegraphics[width=.3\linewidth]{energy_inset.pdf}
\put(-125,108){\textcolor{black}{(a)}} 
\includegraphics[height=.25\linewidth]{bo_tlife.pdf}
\put(-130,110){\textcolor{black}{(b)}} 
\includegraphics[width=0.34\linewidth]{breakup_3d_256_2.pdf}
\put(-130,110){\textcolor{black}{(c)}} 

\includegraphics[width=.32\linewidth]{plots_dir_chap3/Bo}
\put(-130,110){\textcolor{black}{(c)}} 
\includegraphics[width=.32\linewidth]{plots_dir_chap3/vis_life_log_log}
\put(-85,140){\bf (a)}
\includegraphics[width=.32\linewidth]{kinematic_vis.pdf}
\put(-85,140){\bf (a)}
\caption{Plots of (a) the total kinetic energy $E(t)$ and (b) the time derivative 
$dE(t)/dt$ versus $t/\tau_g$ for the 2D CHNS runs {\tt{R16-R20}};
(c) plots versus $Bo$ of the antibubble lifetimes $\tau_1$ and $\tau_2$ calculated, 
respectively, from the energy time series (blue line with circles)
and the minimum antibubble thickness (orange line with squares), for the 
2D CHNS runs {\tt{R1-R16}}.
(d) Plots versus $Bo$ (for $Sc=1.75$), of $\tau_1/\tau_g$ 
for $R_0/h_0=9$ (DNS run {\tt{R13-R16}}, pink
line with triangles), $R_0/h_0=8$ (DNS run {\tt{R17-R20}}, red line with squares), 
$R_0/h_0=9$ (DNS run {\tt{R21-R24}}, blue line with circles); 
(e) log-log plot of $\tau_1/\tau_g$ versus $Sc$, over three orders of magnitude
and at a fixed value of $Bo$; (f) plots of $h_{min}$ versus $t/\tau_g$, at $Bo=1.85\times
10^{-4}$ and $A=0.01$ and $g=99$, for $Sc=1.75$ (blue line with triangles),
	$Sc=66$ (red line with diamonds), and $Sc=245$ (yellow line with crosses).}

\label{sigma_tlife2}
\end{figure*}
}

\REM{
\begin{figure}
\includegraphics[width=.9\linewidth]{3d_s.pdf}

\caption{\label{surface_3d} Time evolution of the antibubble surface area
($S(t)$ for the $3D$ runs ${\tt Q_1, Q_2}$) }

\end{figure}
}

\subsection{Rim-retraction velocity}
\label{subsec:Rim}

When an antibubble ruptures, the surface-tension energy (the interfacial free
energy in the CHNS description) is converted to the kinetic energy of the
antibubble. The rate of change of the former is $\sigma d{\mathcal S}(t)/dt$,
where ${\mathcal S}(t)$ is the outer perimeter of the antibubble in our 2D
DNSs; and the kinetic energy that is released is $E_M=\int_{S_f} \left(\rho
v^{2}({\bm x},t)\right) d{\bm x}$, where $S_f$ is the area of the majority
phase surrounding the minority phase; this causes the film to retract. 
Recall that Figs.~\ref{cont_plots_pierce}(a)-(l) show that, in both our CHNS and VoF 
DNSs, as it ruptures, an antibubble forms a rim, which then retracts.
When this rim retracts, the outer perimeter of
the antibubble reduces by a length $R_f$ on one side after the rupture;
and, if $a$ is the thickness of the vanished film, then
%
%
$E_M  =  \rho v_{rim}^{2} a R_f$, where $v_{rim}$ is the rim-retraction velocity.
%
%
If we assume, furthermore, that the rim retracts with a constant velocity, then
$dv_{rim}/dt=0$, so $dR_f/dt=v_{rim}$.
Finally, we find
\begin{equation}
\sigma d{\mathcal S}(t)/dt  \propto  dE_{M}/dt,~\Rightarrow~v_{rim}  =  \left(\sigma/\rho\right)^{1/2}.
\label{balance}
\end{equation}
This dependence of $v_{rim}$ on $\sigma$ matches the experimental observations
in Ref.~\cite{scheid2012antibubble}. Moreover, this dependence 
is the same in both our CHNS and VoF DNSs, even though, in the former,
antibubble breakup occurs because of gravity-induced thinning, whereas, in the latter, 
this breakup is induced by puncturing the antibubble at its bottom boundary.
The puncturing initial condition has also been used in the
experiments and a theory of antibubble collapse~\cite{sob2015theory,
zou2013collapse}. 

The plot in Fig.~\ref{sigma_tlife}(b) shows that rim velocities, which we obtain from
2D CHNS and VoF DNSs and our 3D axisymmetric VoF runs, for both
high and low $A$, are  in excellent agreement with the theoretical prediction
$v_{th} \sim \sqrt{\sigma/\rho_2 a}$ \cite{sob2015theory}, where $a$ is the
radius of the rim ($\sqrt{h_0 R_0/\pi})$. 

%
%


\subsection{Energy time series}
\label{subsec:EnTime}

%
%

When an antibubble bursts, the surface tension energy is converted into the
kinetic energy of the fluid; this yields a spike in the fluid-energy time
series (see Fig.~\ref{sigma_tlife2}(a)). Therefore, we identify
the breakup time $\tau_1$ as the instant at which $dE(t)/dt$ displays a
maximum. We also define the breakup time $\tau_2$ at which the antibubble-shell
thickness $h_{min}$, at the lower end of the antibubble, vanishes. In
Fig.~\ref{sigma_tlife2}(b) we show plots of both $\tau_1$ and $\tau_2$ versus
$Bo$ for an antibubble with $R_0/h_0=11$ and $\nu=0.967$.  This plot shows that
both our estimates of the collapse times, $\tau_1$ and $\tau_2$, agree with
each other. (For details, see the Supplemental Material~\cite{supmat}.)
%
%

\subsection{Antibubble-breakup times}
\label{subsec:Breakup}

The antibubble-breakup times, which we have defined above, depend on the initial size of 
the antibubble, the surface tension, and the kinematic viscosity. We explore
these dependences below via our 2D CHNS DNSs.

\subsubsection{Dependences on Size and Surface Tension}
\label{subsec:Size}

The size- and surface-tension dependences of the antibubble-breakup time follow
from the plots of $\tau_1/\tau_g$ versus $Bo$, which we give in
Fig.~\ref{sigma_tlife2}(d), for (i) $R_0/h_0=9$ (DNS runs {\tt{R13-R16}}, pink
line with triangles), (ii) $R_0/h_0=8$ (DNS run {\tt{R17-R20}}, red line with
squares), and (iii) $R_0/h_0=9$ (DNS run {\tt{R21-R24}}, blue line with
circles). In all these plots, the viscosity is fixed and the Schmidt number
$Sc=1.75$. Although the scaling of $\tau_1$ by $\tau_g=\sqrt{R/Ag}$ makes
all the curves collapse on top of each, to a large degree, there is a small, but 
noticeable, difference between these curves. 

\REM{
\begin{table}
\resizebox{1.0\linewidth}{!}
{
\begin{tabular}{|l|l|l|l|l|l|l|l|l|}
\hline
& $R_0$ & $R_1$& $h_0$&$A$ & $g$& $\nu$ &$\rho_2$ & $\sigma$\\
\hline
\hline
{\tt {PR1}} & $0.859$& $0.739$  & $0.12$& $0.9$&$1.$& $0.007$ &$1.0$ & $1.66$\\
{\tt {PR2}} & $0.859$& $0.739$  & $0.12$& $0.01$&$100.$&$0.007$ &$1.0$ & $16.7$\\
\hline
\end{tabular}
}

\caption{The parameters  $R_0, R_1, h_0$, $A$, $g$, $\sigma$, $\nu$ and
$\rho_2$ for our 2D (run {\tt {PR1}}) and 3D axisymmetric (run {\tt
{PR2}}) simulations. For 2D and axisymmetric 3D runs we use a square
box of area $L^2=16$ and discretize it with $N^2=2048^2$ collocation
points. Boundary conditions (b.c.): For our 2D simulations we employ
periodic b.c. in all directions, whereas an  axisymmetric b.c. is used
at y=0 and no-slip b.c. ${\bm u}=0$  are used at other boundaries for
our 3D simulation.}
\label{table3_pierce}
\end{table}
}
\REM{
\begin{figure}
\includegraphics[width=.9\linewidth]{kinematic_vis.pdf}
	\caption{\label{kin_visc} Plot of the minimum thickness $h_{min}$ versus scaled time $t/\tau_g$ for different values of the Schimdt number $Sc$  for $R_0/h_0 = 11$. t $Bo = 1.85 \times 10^{-4}$.
 }

\end{figure}
}

\subsubsection{Dependence on the kinematic viscosity}
\label{subsec:Viscosity}

%
%

In
Fig.~\ref{sigma_tlife2}(e), we present a log-log plot of $\tau_1/\tau_g$ versus
$Sc$, over three orders of magnitude and at a fixed value of $Bo$; clearly,
$\tau_1/\tau_g$ increases with $Sc$, with low-$Sc$ and high-$Sc$ asymptotes,
which are consistent with $Sc^{\frac{1}{3}}$ and $Sc^{\frac{2}{3}}$.


In Fig.~\ref{sigma_tlife2}(f), we plot $h_{min}$ versus $t/\tau_g$, at $Bo=1.85\times
10^{-4}$ and $A=0.01$ and $g=99$, for $Sc=1.75$ (blue line with triangles),
$Sc=66$ (red line with diamonds), and $Sc=245$ (yellow line with crosses).  We
note that $h_{min}$ falls rapidly, for $Sc=1.75$, but falls slowly, when
$Sc=5.75$. This observation agrees with experiments~\cite{scheid2012antibubble}.



\subsection{Three-dimensional antibubbles}
\label{subsec:SpTem3D}

\begin{figure*}
\includegraphics[width=.9\linewidth]{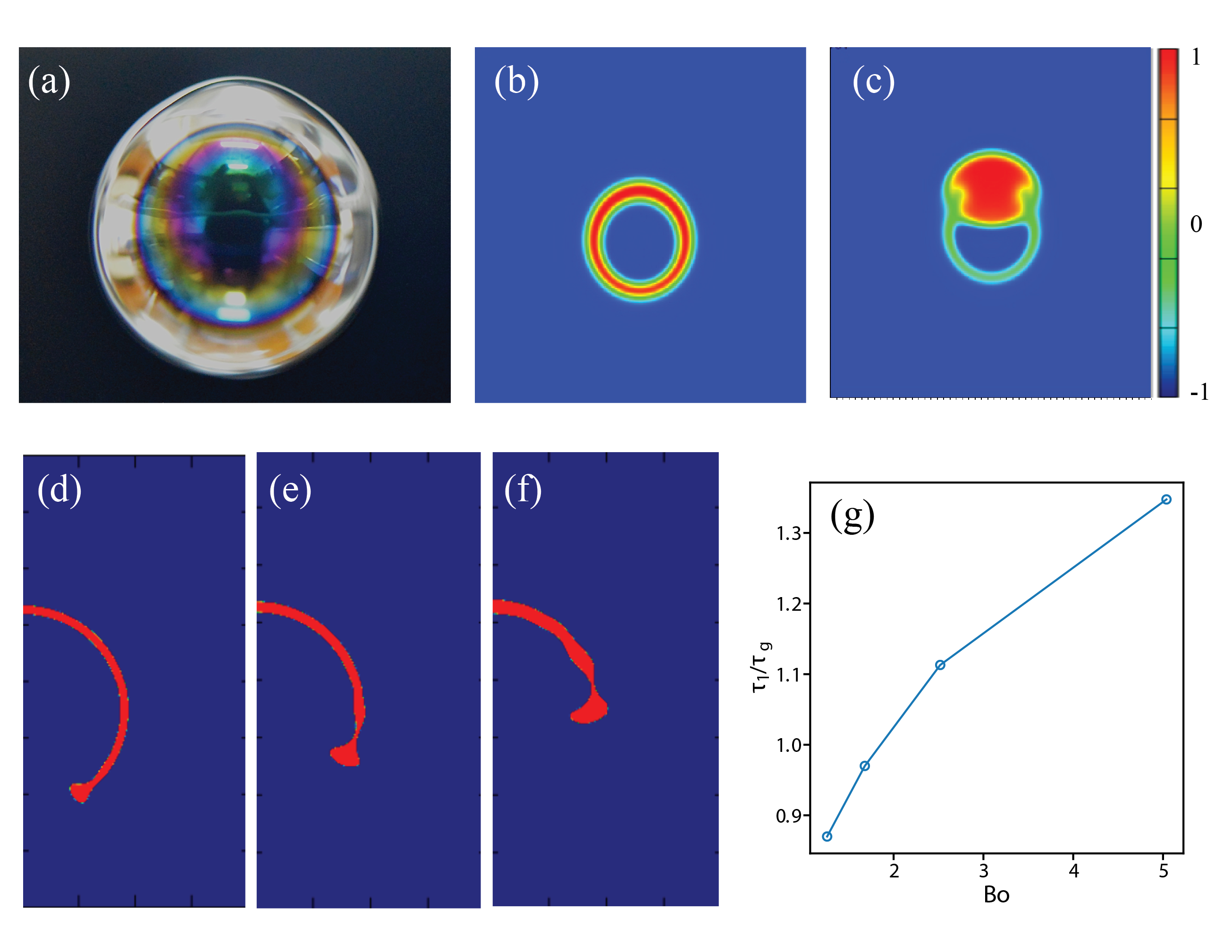}
\caption{(a) Image from an antibubble experiment~\cite{kalelkar2017inveterate}; psuedocolor plot of order parameter showing a section at $y=\pi$ from 3D simulations of antibubbles from run {\tt R2} at (b) $t=0.1\tau_g$ and (c) $t=0.3\tau_g$;
Pseudocolor plots of $\phi$ from our 3D axisymmetric VoF run {\tt{APR7}} run  at  
(d)$t = 0$, (e) $t = 0.1\tau_g$ (f), and $t = 0.3\tau_g$ showing majority (blue) and
minority (red) phases; (g) plot of $\tau_1/\tau_g$ versus $Bo$ from 
our 3D CHNS runs {\tt{Q1-Q6}}.}
\label{exp}
\end{figure*}

We now present illustrative results for 3D antibubbles evolving under gravity
from our CHNS and VoF runs {\tt{Q1-Q7}}, in Table~\ref{table2_mod}, and
{\tt{APR1-APR14}}, in Table~\ref{table3_pierce}, respectively. In Fig.~\ref{exp} (a)
we give, for comparison, an image of an antibubble from an experiment (courtesy of 
C. Kalelkar from Ref.~\cite{kalelkar2017inveterate}). In Figs.~\ref{exp} (b) and (c)
we show psuedocolor plots of two-dimensional sections of $\phi$ (with $y=\pi$ 
in our 3D CHNS run {\tt R2}) at (b) $t=0.1\tau_g$, and (c) $t=0.3\tau_g$. These plots 
show clearly the how the South pole of the antibubble becomes thinner, with the 
passage of time, while a dome develops at its North pole. We also find that, at least
for the parameters we use, the initially spherical antibubble remains axisymmetric.
Therefore, we design our 3D VoF simulations to be axisymmetric. In Figs.~\ref{exp} (d)-(e) we show the rupture of an antibubble in our 3D axisymmetric {\tt{APR7}} VoF run
at  (d)$t = 0$, (e)  $t = 0.1\tau_g$, and (f) $t = 0.3\tau_g$.

%
%
As we did in our 2D CHNS studies, we identify the antibubble-breakup time
$\tau_1$ as the time when $dE/dt$ reaches its maximum value. In
Fig.~\ref{exp}(f) we plot $\tau_1/\tau_g$ versus $Bo$ for our 3D CHNS runs
{\tt{Q1-Q6}}; we find that this plot is qualitatively similar to its 2D CHNS 
counterpart in Fig.~\ref{sigma_tlife2}(b).  We recall that the plot in
Fig.~\ref{sigma_tlife}(b) also shows that rim velocities, which we obtain from
3D axisymmetric VoF runs, are in excellent agreement with the theoretical
prediction~\cite{sob2015theory}.

\REM{
\begin{figure*}
\includegraphics[width=.325\linewidth]{phi07220000.png}
\put(-130,110){\bf \scriptsize \textcolor{white}{(a)}}
\includegraphics[width=.325\linewidth]{phi07220001.png}
\put(-130,110){\bf \scriptsize \textcolor{white}{(b)}}
\includegraphics[width=.325\linewidth]{phi07220002.png}
\put(-130,110){\bf \scriptsize \textcolor{white}{(c)}}

\includegraphics[width=.325\linewidth]{pass07220000.png}
\put(-130,110){\bf \scriptsize(d)}
\includegraphics[width=.325\linewidth]{pass07220000.png}
\put(-130,110){\bf \scriptsize(e)}
\includegraphics[width=.325\linewidth]{pass07220000.png}
\put(-130,110){\bf \scriptsize(f)}

\caption{Top row : Pseudocolor plots of the $\phi$ field with antibubble radius
$R_0/h_0=7$, $\nu=0.0116$ and $\sigma=0.75$ (i.e, $Bo=8.7 \times 10^{-3}$, from
our 3D DNS run {\tt{Q2}}) at (a) $t=0.05\tau_g$, (b)
$t=0.85\tau_g$, (c)
$t=1.25\tau_g$. Bottom row : Pseudocolor plots of the passive scalar field at (d)
$t=0.05\tau_g$, (e) $t=0.85\tau_g$, and (f) $t=1.25\tau_g$.  We show all the values of time
$t$ in multiples of $\tau_g$. (g) Plot of $\tau_1/\tau_g$ versus $Bo$ from our 
3D CHNS runs {\tt{Q1-Q6}}.} 
\label{3d_passive}
\end{figure*}
}

\section{Conclusions}
\label{sec:Conclusions}

An experimental study~\cite{dorbolo2006vita} has the title \textit{Vita brevis
of antibubbles}. We have shown how to use the CHNS system for studying the
spatiotemporal evolution of this short life of antibubbles, in both 2D and 3D.
Our DNSs of the CHNS system allow us to study, numerically and theoretically,
the collapse or breakup of an antibubble because of the gravity-induced
thinning at its South pole. This occurs in experiments via the drainage of air
from the lower end of the antibubble. In some experiments, where antibubbles
are stabilised, say by the addition of surfactants, antibubble rupture is
precipitated by piercing its shell; we have studied such rupture via the
VoF method. By considering Fourier-space spectra, we have shown that the breakup of an
antibubble leads to turbulence, insofar as these spectra have significant
weight over several orders of magnitude of wavenumbers.  Our DNSs have allowed us to
study the dependence of the antibubble lifetime on the surface tension, which
is related inversely to the Bond number $Bo$, the kinematic viscosity, which is
related directly to the Schmidt number $Sc$, and the ratio $R_0/h_0$ of the
initial antibubble radius $R_0$ and its thickness $h_0$. We have also shown how the
antibubble-rim-retraction velocity depends on the surface tension, in both CHNS
and VoF DNSs.  

The dependence of the antibubble lifetime on the surface tension has been
studied earlier~\cite{scheid2012antibubble}, through experiments and numerical
simulation of the balance equations obtained from lubrication theory. A recent
study~\cite{yang2020mathematical} uses the Allen-Cahn-Navier-Stokes equations,
with order-parameter conservation enforced numerically via a Lagrange
multiplier. Our work extends significantly theoretical and DNS studies of
antibubbles, by using the CHNS system, in which order-parameter conservation is
built in manifestly, and the VoF method that can be used fruitfully if
interfaces are very thin and non-Boussinesq effects are present. Our results
agree with earlier results, where they exist. The number of studies on
antibubbles is limited partly because great care has to be exercised to
stabilize antibubbles in terrestrial (as opposed to zero-gravity) experiments.
Often a surfactant has to be introduced into the high-density liquid phase for
such stabilization. We hope our detailed study of the spatiotemporal evolution
of antibubbles, from their initiation to their rupture, and of their effect on
the background fluid, will lead to more experimental studies of the properties
of these ephemeral, but beautiful, inverted bubbles.

\begin{acknowledgments}

We thank Chirag Kalelkar for introducing us to experimental studies of
antibubbles and for the image in  Fig.~\ref{exp} (a) from his
experiments~\cite{kalelkar2017inveterate}. We thank  Nadia Bihari
Padhan and Sumantra Sarkar for discussions. NP thanks Los Alamos
National Laboratory for support; PP and RR acknowledge support from
intramural funds at TIFR Hyderabad from the Department of Atomic Energy
(DAE), India and DST (India) Project No. ECR/2018/001135; RP thanks
CSIR and DST (India) for support and SERC (IISc) for computational
resources.

\end{acknowledgments}


\bibliography{antibubble}

\end{document}